\documentclass[10pt, letterpaper]{article}

\usepackage[margin=1in]{geometry}
\usepackage{amsmath, amsthm, amssymb}
\usepackage{hyperref}
\usepackage{orcidlink}

\usepackage[numbers, sort]{natbib}
\usepackage[labelfont=bf, tableposition=top]{caption}
\usepackage{graphicx, color}

\usepackage[indent=40pt]{parskip}
\title{Quantum walk on simplicial complexes for simplicial community detection}

\author{Euijun Song \orcidlink{0000-0002-5886-4210}\thanks{~TopolyTech, Gyeonggi, Republic of Korea; \href{mailto:drjunsong@gmail.com}{\texttt{drjunsong@gmail.com}}, \href{mailto:esong@topolytech.org}{\texttt{esong@topolytech.org}}}}

\date{\today}

\begin{document}

\maketitle

\begin{abstract}
Quantum walks have emerged as a transformative paradigm in quantum information processing and can be applied to various graph problems. This study explores discrete-time quantum walks on simplicial complexes, a higher-order generalization of graph structures. Simplicial complexes, encoding higher-order interactions through simplices, offer a richer topological representation of complex systems. Since the conventional classical random walk cannot directly detect community structures, we present a quantum walk algorithm to detect higher-order community structures called simplicial communities. We utilize the Fourier coin to produce entangled translation states among adjacent simplices in a simplicial complex. The potential of our quantum algorithm is tested on Zachary's karate club network. This study may contribute to understanding complex systems at the intersection of algebraic topology and quantum walk algorithms.
\end{abstract}

\medskip

\section{Introduction}

Quantum walks have emerged as a powerful paradigm, offering a unique and promising avenue for quantum information processing. Quantum walks, inspired by classical random walks, serve as a cornerstone in the exploration of quantum algorithms and quantum-enhanced computing \cite{Aharonov1993, Travaglione2002}. In contrast to classical random walks, quantum walks show unique behaviors as it produce non-Gaussian probability distributions \cite{Konno2002, Konno2005}. Discrete-time quantum walks represent a quantum counterpart to the discrete-time classical random walks, where a quantum walker traverses lattices or graphs in discrete steps \cite{Aharonov1993}. Discrete-time quantum walks have potential applications in various quantum algorithms for graph problems, including quantum search \cite{Berry2010, Grover1996} and graph community detection \cite{Mukai2020}. Recently, quantum walks have been studied on the generalized graph structures called simplicial complexes \cite{Matsue2016, Matsue2018}, which have much more topological information than graphs or networks.

Network science approaches can extract an interplay between network topology and dynamics of complex systems. While graphs or networks are based on pairwise interactions, higher-order interactions between elements often better capture the topology and dynamics of complex systems \cite{Giusti2016}. Mathematically, higher-order interactions can be represented as simplicial complexes \cite{Hatcher2002, Jonsson2008}, consisting of polytopes called simplices, such as vertices (0-simplices), edges (1-simplices), triangles (2-simplices), etc. Simplicial homology, an algebraic topology tool, can extract rich information about topological features such as high-dimensional holes in simplicial complexes. Using this algebraic topology method, recent studies have explored higher-order dynamics \cite{Millan2020} and homology-based topological data analysis \cite{Song2023, Vipond2021}.

In many real networks, the distribution of edges is globally and locally inhomogeneous. The graph community detection \cite{Girvan2002, Newman2006} has shed light on how edges are locally organized within graphs or networks; that is, graphs are clustered into groups that are densely connected internally, but sparsely connected between distinct groups. The conventional classical random walk cannot directly reveal the graph community structures, because the probability of the classical random walk converges to a stationary distribution that depends only on the degree of a given non-bipartite graph (e.g., see Theorem 7.13 in \cite{Mitzenmacher2005}). Since quantum walks could produce non-stationary probability distributions, quantum algorithms have been developed for detecting the graph community structure \cite{Faccin2014, Mukai2020}. In spectral graph theory, the graph community structure can also be captured by the eigenvectors corresponding to non-zero eigenvalues of the graph Laplacian \cite{Newman2013}. Krishnagopal and Bianconi \cite{Krishnagopal2021} generalized the graph community concept to simplicial complexes and revealed that simplicial communities can be identified from the support of non-zero eigenvectors of the higher-order Laplacian, also known as the Hodge Laplacian. This novel concept of the higher-order community structure has been poorly investigated.

In this study, we present discrete-time quantum walk on simplicial complexes for detecting simplicial communities, as a higher-order generalization of the method proposed by Mukai and Hatano \cite{Mukai2020}. In Section \ref{secbackground}, we review the fundamentals of algebraic topology, such as simplicial complexes and higher-order Laplacian. In Section \ref{secsimsomm}, the simplicial community and modularity are mathematically defined, and the symmetry between up-communities and down-communities is presented. In Section \ref{secqw}, we present the discrete-time quantum walk method for detecting simplicial communities, based on the Fourier quantum walk \cite{Mackay2002, Saito2019}. The long-time average of the transition probability is theoretically calculated to identify simplicial communities. Finally, in Section \ref{secrealnet}, we test our quantum algorithm in real networks.

\section{Simplicial complex and higher-order Laplacian} \label{secbackground}
In this section, we briefly review the key algebraic topology background, including the concept of simplicial complexes \cite{Hatcher2002, Jonsson2008} and the Hodge Laplacian \cite{Lim2020}.

\subsection{Simplicial complex}
A simplicial complex is a type of higher-order network encoding information about higher-order interactions. A simplicial complex consists of a set of polytopes called simplices. An \textit{$n$-dimensional simplex ($n$-simplex)} $\sigma$ is defined as a set of $n+1$ nodes ($n\geq 0$) with an assigned orientation:
\begin{equation}
\sigma=(v_0, v_1, \ldots, v_n).
\end{equation}
An \textit{$n'$-face} of an $n$-simplex $\sigma$ ($n' < n$) is an $n'$-simplex generated by a subset of the nodes of $\sigma$. A \textit{simplicial complex} $\mathcal{K}$ is then defined as a collection of simplices closed under the inclusion of the faces of simplices. A simplicial \textit{$n$-chain} $c_n$ is a finite linear combination of $n$-simplices in a simplicial complex $\mathcal{K}$, as follows:
\begin{equation}
c_n=\sum_{i}{b_{i}\sigma_{i}}
\end{equation}
where each $b_i$ is an integer coefficient, or generally an element of any field. A set of $n$-chains on $\mathcal{K}$ forms the free abelian group $C_n$. For $n>0$, the \textit{boundary operator} $\partial_{n}: C_{n}\to C_{n-1}$ is defined as the homomorphism that maps $n$-chains to $(n-1)$-chains \cite{Hatcher2002, Jonsson2008}:
\begin{equation}
(v_0, v_1, \ldots, v_n) \mapsto \sum_{i=0}^{n}{(-1)^{i}(v_0,\ldots,v_{i-1},v_{i+1},\ldots,v_n)}.
\end{equation}
One fundamental property of the boundary operator is that the boundary of the boundary is zero, i.e., $\partial_{n}\partial_{n+1}=0$ (see Lemma 2.1 in Hatcher \cite{Hatcher2002}). Let $N_n$ be the number of $n$-simplices in a simplicial complex $\mathcal{K}$. Using a basis for simplices of $\mathcal{K}$, the boundary operator $\partial_n$ can be represented as an incidence matrix $B_n\in\mathbb{R}^{N_{n-1}\times N_{n}}$, as follows \cite{Baccini2022}:
\begin{equation}
B_n(\sigma_{\sim k}, \sigma)=(-1)^k
\end{equation}
where $\sigma=(v_0,v_1,\ldots,v_n)$ and $\sigma_{\sim k}=(v_0,\ldots,v_{k-1},v_{k+1},\ldots,v_n)$. Since the boundary of the boundary is zero, $B_{n}B_{n+1}=0$ holds for $n>0$.

\subsection{Higher-order Laplacian and adjacency}
In graphs or networks, the graph Laplacian is defined as $L=D-A$, where $D$ is the degree matrix and $A$ is the adjacency matrix of a graph. Using an incidence matrix $B_1$, the graph Laplacian can be written as $L=B_{1}B_{1}^{T}$. The graph Laplacian is known to be central to the diffusion process on a graph. The graph Laplacian was generalized to simplicial complexes by Eckmann \cite{Eckmann1944}. The higher-order Laplacian, also known as the Hodge Laplacian \cite{Lim2020}, encodes the higher-order topology and can be used to examine high-dimensional diffusion on a simplicial complex.

We first define higher-order adjacency for a simplicial complex, as previously studied \cite{Estrada2018, Hernandez2020, Muhammad2006}. Two $n$-simplices $\sigma$ and $\sigma'$ ($n\geq 0$) are \textit{upper adjacent} if there exists an $(n+1)$-simplex $\gamma$ such that $\sigma\subset\gamma$ and $\sigma'\subset\gamma$; that is, they are both faces of a common $(n+1)$-simplex. This upper adjacency is denoted by $\sigma\frown\sigma'$. Similarly, two $n$-simplices $\sigma$ and $\sigma'$ ($n>0$) are \textit{lower adjacent} if there exists an $(n-1)$-simplex $\gamma$ such that $\gamma\subset\sigma$ and $\gamma\subset\sigma'$; that is they share a common $(n-1)$-face. This lower adjacency is denoted by $\sigma\smile\sigma'$. We also define the upper degree and lower degree on a simplicial complex. For an $n$-simplex $\sigma$, the \textit{upper degree} $d^{u}_{n}(\sigma)$ is the number of $(n+1)$-simplices of which $\sigma$ is a face. The \textit{lower degree} $d^{l}_{n}(\sigma)$ is the number of $(n-1)$-faces in $\sigma$. It is easy to check $d^{l}_{n}(\sigma)=n+1$ for $n>0$. The upper and lower adjacency matrices $A^{u}_{n}, A^{l}_{n}\in\mathbb{R}^{N_{n}\times N_{n}}$ for $n$-simplices are defined as follows \cite{Estrada2018}:
 \begin{equation}
A^{u}_{n}(\sigma,\sigma') = \begin{cases}
1 & \text{if } \sigma\frown\sigma' \\
0 & \text{otherwise}
\end{cases}
\end{equation}
and
 \begin{equation}
A^{l}_{n}(\sigma,\sigma') = \begin{cases}
1 & \text{if } \sigma\smile\sigma' \\
0 & \text{otherwise.}
\end{cases}
\end{equation}

The Hodge Laplacian describes higher-dimensional diffusion on a simplicial complex. The diffusion process on a simplicial complex can be performed through either upper adjacent simplices or lower adjacent simplices. Formally, the \textit{Hodge Laplacian} $L_n\in\mathbb{R}^{N_{n}\times N_{n}}$ on a simplicial complex $\mathcal{K}$ is defined as follows \cite{Baccini2022, Krishnagopal2021, Lim2020}:
\begin{equation}
L_n = \begin{cases}
L^{up}_{0} & \text{for } n=0 \\
L^{up}_{n}+L^{down}_{n} & \text{for } n>0
\end{cases}
\end{equation}
where $L^{up}_{n}=B_{n+1}B_{n+1}^{T}$ and $L^{down}_{n}=B_{n}^{T}B_{n}$ represent diffusion through upper adjacent simplicies and lower adjacent simplices, respectively. For $n>0$ and $n$-simplices $\sigma, \sigma'\in\mathcal{K}$, the matrix elements of $L^{up}_{n}$ and $L^{down}_{n}$ can be written as follows \cite{Muhammad2006}:
\begin{equation}
L^{up}_{n}(\sigma,\sigma') = \begin{cases}
d^{u}_{n}(\sigma) & \text{if } \sigma=\sigma' \\
-1 & \text{if } \sigma\frown\sigma',~\Omega(\sigma)=\Omega(\sigma') \\
1 & \text{if } \sigma\frown\sigma',~\Omega(\sigma)\neq\Omega(\sigma') \\
0 & \text{otherwise}
\end{cases}
\end{equation}
and
\begin{equation}
L^{down}_{n}(\sigma,\sigma') = \begin{cases}
d^{l}_{n}(\sigma) = n+1 & \text{if } \sigma=\sigma' \\
1 & \text{if } \sigma\smile\sigma',~\Omega(\sigma)=\Omega(\sigma') \\
-1 & \text{if } \sigma\smile\sigma',~\Omega(\sigma)\neq\Omega(\sigma') \\
0 & \text{otherwise}
\end{cases}
\end{equation}
where $\Omega(\sigma)$ indicates the orientation of $\sigma$. If two $n$-simplices are upper adjacent, then they are also lower adjacent. Therefore, the off-diagonal element of the Hodge Laplacian $L_n$ is non-zero if and only if $\neg(\sigma\frown\sigma')$ and $\sigma\smile\sigma'$. Based on this property of the Hodge Laplacian, we say that two $n$-simplices $\sigma, \sigma'\in\mathcal{K}$ are \textit{adjacent} in $\mathcal{K}$ if $\neg(\sigma\frown\sigma')$ and $\sigma\smile\sigma'$.

The Hodge Laplacians have useful algebraic properties \cite{Baccini2022, Krishnagopal2021}. By the definition of the Hodge Laplacians, we obtain $L^{up}_{n}L^{down}_{n}=0$ and $L^{down}_{n}L^{up}_{n}=0$. This implies that
\begin{eqnarray}
\textrm{im}(L^{up}_{n}) &\subseteq& \ker(L^{down}_{n}) \nonumber \\
\textrm{im}(L^{down}_{n}) &\subseteq& \ker(L^{up}_{n}).
\end{eqnarray}
In addition, the operator $\mathcal{L}_{n}=\partial_{n+1}\partial_{n+1}^{*}+\partial_{n}^{*}\partial_{n}$ corresponding to the Hodge Laplacian matrix $L_{n}$ has the following property called the \textit{Hodge decomposition} \cite{Lim2020}:
\begin{equation}
C_{n}=\textrm{im}(\partial_{n}^{*})\oplus\ker(\mathcal{L}_{n})\oplus\textrm{im}(\partial_{n+1}).
\end{equation}
It is known that $\ker(\mathcal{L}_{n})$ is isomorphic to the $n$th homology group $H_n$, implying that the multiplicity of the zero eigenvalues of the Hodge Laplacian is the same as the number of $n$-dimensional holes in a simplicial complex.

\section{Simplicial community} \label{secsimsomm}

\subsection{Walks on a simplicial complex}
Graph community structures can be generalized to determine higher-order community structures on a simplicial complex. We first extend the concept of walks to a simplicial complex \cite{Hernandez2020}. Two $n$-simplices $\sigma, \sigma'\in\mathcal{K}$ are said to be \textit{$n$-upper-connected} ($n\geq 0$) if there exists a sequence of $n$-simplices $\sigma=s_0, s_1, \ldots, s_{p-1}, s_{p}=\sigma'$ such that any two consecutive simplices are upper adjacent, i.e., $s_{i}\frown s_{i+1}$ for all $0\leq i\leq p-1$. We call this sequence of simplices an \textit{$n$-upper-walk}. Similarly, two $n$-simplices $\sigma, \sigma'\in\mathcal{K}$ are said to be \textit{$n$-lower-connected} ($n>0$) if there exists a sequence of $n$-simplices $\sigma=s_0, s_1, \ldots, s_{p-1}, s_{p}=\sigma'$ such that any two consecutive simplices are lower adjacent, i.e., $s_{i}\smile s_{i+1}$ for all $0\leq i\leq p-1$. We call this sequence of simplices an \textit{$n$-lower-walk}.

\subsection{Higher-order community structure}
As previously studied by Krishnagopal and Bianconi \cite{Krishnagopal2021}, a simplicial community can be defined as a set of $n$-simplicies that are $n$-connected. Specifically, an \textit{$n$-up community} is a set of $n$-simplicies that are $n$-upper-connected, and an \textit{$n$-down community} is a set of $n$-simplicies that are $n$-lower-connected. We present one important property of the simplicial community---the symmetry between $n$-up communities and $(n+1)$-down communities.

Let us consider that $(n+1)$-simplices of a simplicial complex $\mathcal{K}$ are partitioned into $(n+1)$-down communities
\begin{equation}
\Pi_{n+1}^{d}=\left\{\pi_1^d,\ldots,\pi_{g_{n+1}}^{d}\right\}
\end{equation}
where $\pi_i^d$ is the set of $(n+1)$-simplices in the $i$-th $(n+1)$-down community. Ideally, any two $(n+1)$-simplices belonging to distinct $(n+1)$-down communities are not $(n+1)$-lower-connected, and $\pi_i^d \cap \pi_j^d=\emptyset$ ($i\neq j$).
In a similar way, consider $n$-up communities of a simplicial complex $\mathcal{K}$:
\begin{equation}
\Pi_{n}^{u}=\left\{\pi_1^u,\ldots,\pi_{h_{n}}^{u}\right\}.
\end{equation}
We define $\hat{\Pi}_{n}^{u}=\left\{\pi_i^u \mid |\pi_i^u|>1,~\pi_i^u\in\Pi_{n}^{u} \right\}$ by excluding the isolated $n$-simplices.
Then, $(n+1)$-down communities $\Pi_{n+1}^{d}$ is isomorphic to the $n$-up communities (excluding the isolated $n$-simplices) $\hat{\Pi}_{n}^{u}$. We can prove this symmetry property by defining the isomorphism $\zeta: \Pi_{n+1}^{d}\to \hat{\Pi}_{n}^{u}$ as follows:
\begin{equation}
\pi_i^d \mapsto \left\{ \sigma_{n}'~\middle|~\sigma_{n}' \textrm{ is an } n\textrm{-face of } \sigma_{n+1}\in\pi_i^d \right\}.
\end{equation}
We can also derive this property from the fact that simplicial communities can be captured by the support of the eigenvectors associated with non-zero eigenvalues of $L^{up}_{n}=B_{n+1}B_{n+1}^{T}$ (or $L^{down}_{n+1}=B_{n+1}^{T}B_{n+1}$) \cite{Krishnagopal2021}.

Based on the symmetry between up-communities and down-communities, we only consider $n$-down simplicial communities ($n>0$) in the rest of the present work. The term simplicial community refers to the $n$-down simplicial community.

\subsection{Simplicial modularity}
Ideally, any two $n$-simplices belonging to distinct simplicial communities are not $n$-connected. However, in many complex networks, it is often necessary to detect community structures that are densely connected internally but sparsely connected between distinct communities. In graphs or networks, algorithms have been developed for maximizing modularity, which is defined as the difference between the fraction of the edges within the given communities and the expected fraction at random distribution \cite{Newman2006}. We can extend the concept of modularity to the $n$-down simplicial communities $\Pi_{n}^{d}=\left\{\pi_1^d,\ldots,\pi_{g_{n}}^{d}\right\}$. Define a matrix $W_{n}\in \mathbb{R}^{N_{n}\times |\Pi_{n}^{d}|}$, by $W_{n}(\sigma, \pi_i^d)$ to be 1 if $\sigma\in \pi_i^d$ and otherwise zero. We denote the \textit{lower neighborhood}, a set of lower-adjacent simplices of an $n$-simplex $\sigma$, by
\begin{equation}
\mathcal{N}^{l}(\sigma)=\left\{\sigma'~\middle|~\sigma\smile\sigma' \right\}.
\end{equation}
We then define the \textit{simplicial modularity} for the $n$-down simplicial communities as follows:
\begin{equation} \label{simmodularity}
Q_n=\frac{1}{m_n}\textrm{Tr}\left(W^{T}_{n}M_{n}W_{n}\right)
\end{equation}
where $m_n=\sum_{\sigma} \left|\mathcal{N}^{l}(\sigma)\right|$ and $M_{n}\in\mathbb{R}^{N_{n}\times N_{n}}$ is the modularity matrix:
\begin{equation}
M_{n}(\sigma,\sigma') = A^{l}_{n}(\sigma,\sigma') - \frac{\left|\mathcal{N}^{l}(\sigma)\right|\cdot \left|\mathcal{N}^{l}(\sigma')\right|}{m_n}.
\end{equation}
Although modularity maximization is one of the most popular approaches for detecting community structures in graphs, it has a resolution limit leading to merging small clusters and it has to determine the community number parameter. In the next section, we present the quantum walk algorithm for detecting simplicial communities.

\section{Quantum walk for simplicial community detection} \label{secqw}

\subsection{Discrete-time quantum walk on simplicial complexes}
Since the probability of the classical random walk converges to a stationary distribution \cite{Mitzenmacher2005}, the conventional classical random walk is unlikely to directly detect community structures in graphs and simplicial complexes. Here, we present the discrete-time quantum walk on simplicial complexes for detecting $n$-down simplicial communities ($n>0$). We directly generalize the community detection method proposed by Mukai and Hatano \cite{Mukai2020} to the simplicial community. The method utilizes the \textit{Fourier coin} (or the Fourier quantum walk), which has been studied in high-dimensional lattices \cite{Mackay2002}, regular graphs \cite{Saito2019}, and complex networks \cite{Mukai2020}.

Let $\sigma_1, \sigma_2, \ldots, \sigma_{N_n}\in\mathcal{K}$ be $n$-simplices in a simplicial complex $\mathcal{K}$. The total Hilbert space is given by $\mathcal{H}=\bigoplus_{i=1}^{N_n} \mathcal{H}_{\sigma_i}$, consisting of the Hilbert space corresponding to each $\sigma_i$. Each Hilbert space $\mathcal{H}_{\sigma_i}$ is spanned by the orthonormal basis $\left\{\vert\sigma_i\to\sigma_j\rangle \mid \sigma_j\in\mathcal{N}^{l}(\sigma_i)\right\}$, where $\vert\sigma_i\to\sigma_j\rangle$ represents the translation state jumping from $\sigma_i$ to its lower-adjacent simplex $\sigma_j\in \mathcal{N}^{l}(\sigma_i)$. The dimension of the total Hilbert space $\mathcal{H}$ is equal to $m_n=\sum_{\sigma_i} \left|\mathcal{N}^{l}(\sigma_i)\right|$. The quantum state $\vert\Psi(t)\rangle\in\mathcal{H}$ is given by
\begin{eqnarray}
\vert\Psi(t)\rangle &=& \sum_{\sigma_i\smile\sigma_j} \psi_{\sigma_i,\sigma_j}(t) \vert\sigma_i\to\sigma_j\rangle \nonumber \\
&=& \sum_{i=1}^{N_n} \sum_{\sigma_j\in\mathcal{N}^{l}(\sigma_i)} \psi_{\sigma_i,\sigma_j}(t) \vert\sigma_i\to\sigma_j\rangle.
\end{eqnarray}
The sum of probabilities is equal to 1, as $\langle\Psi(t)\vert\Psi(t)\rangle=\sum_{\sigma_i\smile\sigma_j} \left|\psi_{\sigma_i,\sigma_j}(t)\right|^2=1$.

We now define \textit{unitary time evolution operator} $U: \mathcal{H}\to\mathcal{H}$ to specify how our quantum walk works. The time evolution of the quantum state is given by $\vert\Psi(t)\rangle=U^{t}\vert\Psi(0)\rangle$. The unitary operator is basically given by $U=SC$, where $S$ is the shift operator and $C$ is the coin operator. We define the \textit{coin operator} $C: \mathcal{H}\to\mathcal{H}$ as a direct sum of the \textit{Fourier quantum walks} \cite{Mackay2002, Saito2019} $F_{\sigma_i}: \mathcal{H}_{\sigma_i}\to \mathcal{H}_{\sigma_i}$ as follows:
\begin{equation}
C = \bigoplus_{i=1}^{N_n} F_{\sigma_i}
\end{equation}
and
\begin{equation}
F_{\sigma_i} = \frac{1}{\sqrt{\left|\mathcal{N}^{l}(\sigma_i)\right|}} \sum_{\substack{\sigma_{j_\alpha} \in\mathcal{N}^{l}(\sigma_i) \\ \alpha=0,1,\ldots,|\mathcal{N}^{l}(\sigma_i)|-1}}~\sum_{\substack{\sigma_{j_\beta} \in\mathcal{N}^{l}(\sigma_i) \\ \beta=0,1,\ldots,|\mathcal{N}^{l}(\sigma_i)|-1}} \omega^{\alpha\beta} \vert\sigma_i\to\sigma_{j_\beta}\rangle \langle\sigma_i\to\sigma_{j_\alpha}\vert
\end{equation}
where $\omega=e^{\frac{2\pi\sqrt{-1}}{|\mathcal{N}^{l}(\sigma_i)|}}$. This Fourier coin operator $F_{\sigma_i}$ produces entanglement by transforming any translation state $\vert\sigma_i\to\sigma_{j_\alpha}\rangle$ into an equally-weighted superposition of all the translation states $\vert\sigma_i\to\sigma_{j_\beta}\rangle$ based on $\sigma_i$. We define the \textit{shift operator} $S: \mathcal{H}\to\mathcal{H}$ as follows:
\begin{equation}
S = \sum_{\sigma_i\smile\sigma_j} \vert\sigma_j\to\sigma_i\rangle \langle\sigma_i\to\sigma_j\vert.
\end{equation}
For lower-adjacent simplices $\sigma_i$ and $\sigma_j$, this shift operator $S$ maps $\vert\sigma_i\to\sigma_j\rangle$ to $\vert\sigma_j\to\sigma_i\rangle$, i.e., moving the walker from $\sigma_i$ to $\sigma_j\in \mathcal{N}^{l}(\sigma_i)$, and vice versa. It can be represented as a direct sum of $2\times2$ permutation matrices:
\begin{equation}
S = \bigoplus_{\substack{\sigma_i\smile\sigma_j \\ i<j}}
\begin{bmatrix}
0 & 1 \\
1 & 0
\end{bmatrix}.
\end{equation}

\subsection{Transition probability}
We calculate the transition probability from one $n$-simplex $\sigma_x$ to another $n$-simplex $\sigma_y$ (not necessarily lower adjacent). We set the initial quantum state $\vert\Psi(0)\rangle$ such that the quantum walk starts from $\vert\sigma_x\to\sigma_v\rangle$ ($\sigma_v\in \mathcal{N}^{l}(\sigma_x)$) and calculate the average probability over $\mathcal{N}^{l}(\sigma_x)$. The \textit{normalized transition probability} that the quantum walker starting from $\sigma_x$ reaches $\sigma_y$ at time $t$ is given by
\begin{equation} \label{transitionprob1}
p(\sigma_x\to\sigma_y; t) = \frac{1}{\left|\mathcal{N}^{l}(\sigma_x)\right| \cdot \left|\mathcal{N}^{l}(\sigma_y)\right|} \sum_{\sigma_w\in \mathcal{N}^{l}(\sigma_y)} \sum_{\sigma_v\in \mathcal{N}^{l}(\sigma_x)} \left| \langle\sigma_y\to\sigma_w \vert U^{t} \vert \sigma_x\to\sigma_v\rangle \right|^2
\end{equation}
where $\vert\sigma_x\to\sigma_v\rangle$ is the initial state and $\vert\sigma_y\to\sigma_w\rangle$ is the final state at time $t$.

Let us calculate the \textit{long-time average} of the normalized transition probability:
\begin{equation} \label{longtimeprob}
q(\sigma_x\to\sigma_y) = \lim_{T\to\infty} \frac{1}{T} \sum_{t=1}^{T} p(\sigma_x\to\sigma_y; t).
\end{equation}
In the practical situation, $q(\sigma_x\to\sigma_y)$ can be approximated by finite-time results for the quantum walk. Here, we theoretically calculate $q(\sigma_x\to\sigma_y)$ using the spectral decomposition of the unitary operator $U$. Let $e^{i\theta_k}$ ($i=\sqrt{-1}$) and $\vert\Phi_k\rangle$ be the eigenvalue and the corresponding eigenstate of the unitary operator $U$, respectively ($\theta_k\in [0, 2\pi)$, $k=1,2,\ldots,m_n$). Suppose the eigenvalues of $U$ are non-degenerate. The spectral decomposition of $U$ is given by $U=\sum_{k=1}^{m_n} \vert\Phi_k\rangle e^{i\theta_k} \langle\Phi_k\vert$.
Hence,
\begin{eqnarray} \label{transitionprob2}
& & \lim_{T\to\infty} \frac{1}{T} \sum_{t=1}^{T} \left| \langle\sigma_y\to\sigma_w \vert U^{t} \vert \sigma_x\to\sigma_v\rangle \right|^2 \nonumber \\
& & = \lim_{T\to\infty} \frac{1}{T} \sum_{t=1}^{T} \left| \sum_{k=1}^{m_n} e^{i\theta_{k}t} \langle\sigma_y\to\sigma_w \vert \Phi_k\rangle \langle\Phi_k \vert \sigma_x\to\sigma_v\rangle \right|^2 \nonumber \\
& & = \lim_{T\to\infty} \frac{1}{T} \sum_{t=1}^{T} \sum_{k_1,k_2} e^{i(\theta_{k_1}-\theta_{k_2})t} \langle\sigma_y\to\sigma_w \vert \Phi_{k_1}\rangle \langle\Phi_{k_1} \vert \sigma_x\to\sigma_v\rangle \langle\sigma_x\to\sigma_v \vert \Phi_{k_2}\rangle \langle\Phi_{k_2} \vert \sigma_y\to\sigma_w\rangle \nonumber \\
& & = \sum_{k_1,k_2} \delta_{\theta_{k_1}, \theta_{k_2}} \langle\sigma_y\to\sigma_w \vert \Phi_{k_1}\rangle \langle\Phi_{k_1} \vert \sigma_x\to\sigma_v\rangle \langle\sigma_x\to\sigma_v \vert \Phi_{k_2}\rangle \langle\Phi_{k_2} \vert \sigma_y\to\sigma_w\rangle \nonumber \\
& & = \sum_{k=1}^{m_n} \left| \langle\sigma_y\to\sigma_w \vert \Phi_{k}\rangle \langle\Phi_{k} \vert \sigma_x\to\sigma_v\rangle \right|^2
\end{eqnarray}
where $\delta_{\theta_{k_1}, \theta_{k_2}}$ is the Kronecker delta, and we used the following property (see Appendix \ref{appendix1} for proof):
\begin{equation} \label{longtimeeigen}
\lim_{T\to\infty} \frac{1}{T} \sum_{t=1}^{T} e^{i(\theta_{k_1}-\theta_{k_2})t} = \delta_{\theta_{k_1}, \theta_{k_2}}.
\end{equation}
By combining Eq. \ref{transitionprob1} and Eq. \ref{transitionprob2}, we have
\begin{equation} \label{probtheory}
q(\sigma_x\to\sigma_y) = \frac{1}{\left|\mathcal{N}^{l}(\sigma_x)\right| \cdot \left|\mathcal{N}^{l}(\sigma_y)\right|} \sum_{\sigma_w\in \mathcal{N}^{l}(\sigma_y)} \sum_{\sigma_v\in \mathcal{N}^{l}(\sigma_x)} \sum_{k=1}^{m_n} \left| \langle\sigma_y\to\sigma_w \vert \Phi_{k}\rangle \langle\Phi_{k} \vert \sigma_x\to\sigma_v\rangle \right|^2.
\end{equation}

We can estimate the lower bound of $q(\sigma_x\to\sigma_y)$ by the average transition amplitude, as follows:
\begin{eqnarray}
q(\sigma_x\to\sigma_y) &=& \sum_{k=1}^{m_n} \frac{1}{\left|\mathcal{N}^{l}(\sigma_x)\right| \cdot \left|\mathcal{N}^{l}(\sigma_y)\right|} \sum_{\sigma_w\in \mathcal{N}^{l}(\sigma_y)} \sum_{\sigma_v\in \mathcal{N}^{l}(\sigma_x)} \left| \langle\sigma_y\to\sigma_w \vert \Phi_{k}\rangle \langle\Phi_{k} \vert \sigma_x\to\sigma_v\rangle \right|^2 \nonumber \\
&\geq& \sum_{k=1}^{m_n} \left| \frac{1}{\left|\mathcal{N}^{l}(\sigma_x)\right| \cdot \left|\mathcal{N}^{l}(\sigma_y)\right|} \sum_{\sigma_w\in \mathcal{N}^{l}(\sigma_y)} \sum_{\sigma_v\in \mathcal{N}^{l}(\sigma_x)} \langle\sigma_y\to\sigma_w \vert \Phi_{k}\rangle \langle\Phi_{k} \vert \sigma_x\to\sigma_v\rangle \right|^2 \nonumber \\
&=& \sum_{k=1}^{m_n} \left| \langle\sigma_y\vert \Phi_{k}\rangle \langle\Phi_{k} \vert\sigma_x\rangle \right|^2
\end{eqnarray}
where $\vert\sigma_x\rangle$ and $\vert\sigma_y\rangle$ are equally-weighted superpositions of all the translation states based on $\sigma_x$ and $\sigma_y$, respectively:
\begin{eqnarray}
\vert\sigma_x\rangle &=& \frac{1}{\left|\mathcal{N}^{l}(\sigma_x)\right|} \sum_{\sigma_v\in \mathcal{N}^{l}(\sigma_x)} \vert\sigma_x\to\sigma_v\rangle \nonumber \\
\vert\sigma_y\rangle &=& \frac{1}{\left|\mathcal{N}^{l}(\sigma_y)\right|} \sum_{\sigma_w\in \mathcal{N}^{l}(\sigma_y)} \vert\sigma_y\to\sigma_w\rangle.
\end{eqnarray}

\subsection{Simplicial community detection}
Since the Fourier quantum walk on graphs is likely to be localized in a community containing the initial node \cite{Mukai2020}, we expect our generalized framework could work for detecting simplicial communities. We present the quantum walk algorithm for detecting simplicial communities based on the long-time average of the normalized transition probability $q(\sigma_x\to\sigma_y)$ (Eq. \ref{longtimeprob}).
\begin{enumerate}
\item Set a starting $n$-simplex $\sigma_x\in\mathcal{K}$ whose $\left|\mathcal{N}^{l}(\sigma_x)\right|$ is the maximum over all the unassigned $n$-simplices.
\item Compute $q(\sigma_x\to\sigma_y)$ for each of the unassigned $n$-simplices $\sigma_y\in\mathcal{K}$.
\item Assign an $n$-simplex $\sigma_y$ to a member of the simplicial community of $\sigma_x$, if
\begin{equation} \label{probcriteria}
q(\sigma_x\to\sigma_y)>\frac{1}{m_n}.
\end{equation}
\item Repeat until all the $n$-simplices are assigned to the simplicial communities.
\end{enumerate}
The threshold $1/{m_n}$ (Eq. \ref{probcriteria}) is set to be the stationary probability of the classical random walk on regular graphs, so that the transition probability that the quantum walker reaches the same simplicial community of the starting simplex is higher than that expected by chance.

We can theoretically calculate the long-time average $q(\sigma_x\to\sigma_y)$ using the spectrum of the unitary operator $U$, as shown in Eq. \ref{probtheory}. Since the spectral decomposition requires a large computational complexity of $O(\textrm{poly } m_n)$, the finite-time approximation by repeatedly applying the unitary operator $U$ is a more efficient and natural way in quantum computation. We therefore approximate $q(\sigma_x\to\sigma_y)$ by the finite-time results for the quantum walk. In our simulation analysis, we compute the finite-time average of the normalized transition probability over 100 time steps from $t=1$ to $t=100$, and denote this finite-time approximation by $\tilde{q}_{T}(\sigma_x\to\sigma_y)$ (e.g., $T=100$):
\begin{equation}
\tilde{q}_{T}(\sigma_x\to\sigma_y) = \frac{1}{T} \sum_{t=1}^{T} p(\sigma_x\to\sigma_y; t).
\end{equation}

The probability distributions of discrete-time quantum walks on simple graphs like Zachary's karate club network \cite{Zachary1977} has been comprehensively studied by Mukai and Hatano \cite{Mukai2020}.  Here, we further show that the error of this finite-time approximation converges to 0 as $T\to\infty$, with the order of $O(1/T)$ (see Appendix \ref{appendix2} for proof):
\begin{equation} \label{errorterm}
\left| \tilde{q}_{T}(\sigma_x\to\sigma_y) - q(\sigma_x\to\sigma_y) \right| = O\left(\frac{1}{T}\right).
\end{equation}

\section{Application to real networks} \label{secrealnet}

\begin{figure}[h]
\centering
\includegraphics[width=0.7\textwidth]{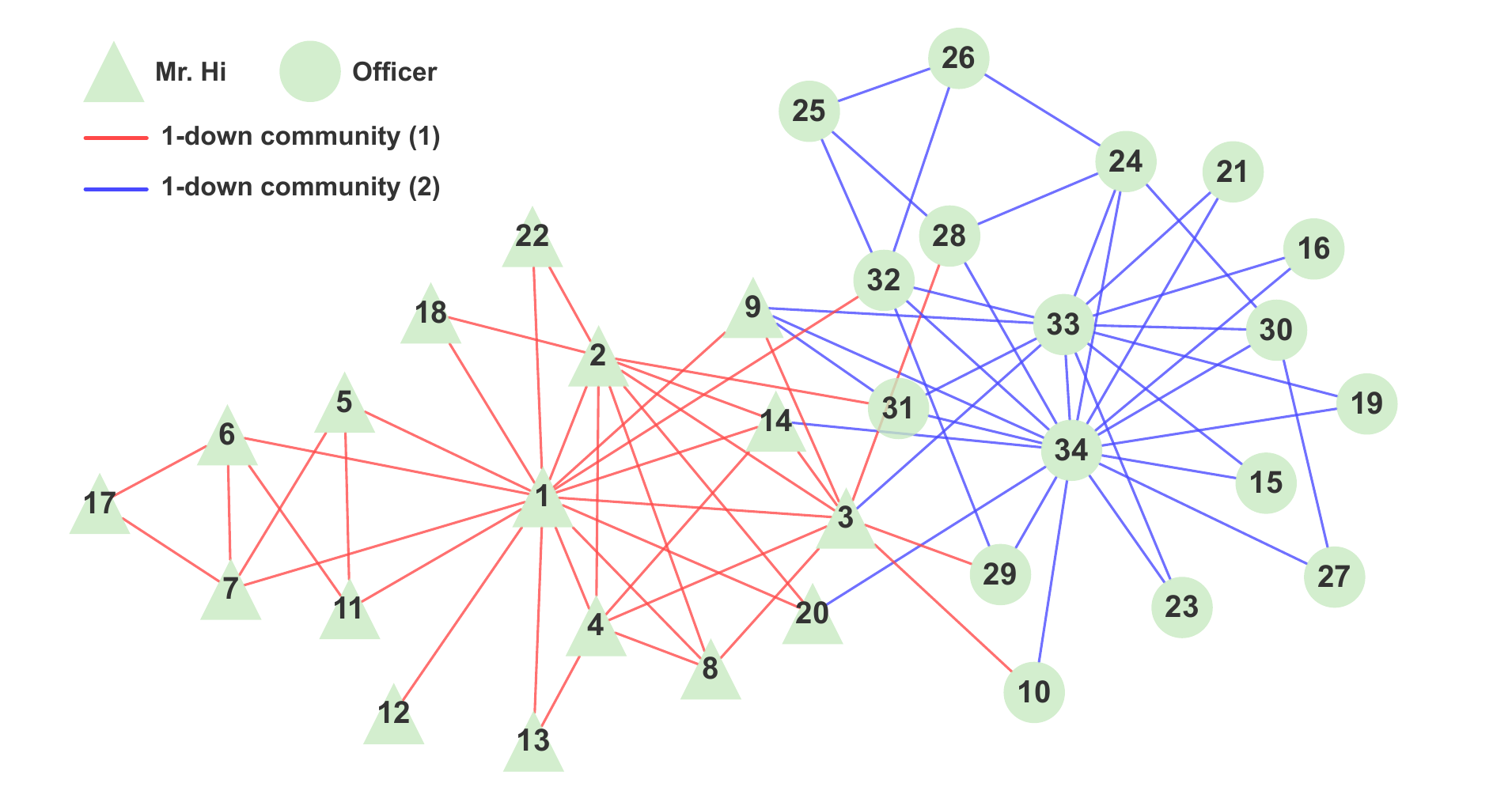}
\caption{The 1-down communities of Zachary's karate club network.}
\label{fig:1}
\end{figure}

We test our quantum walk algorithm to Zachary's karate club network \cite{Zachary1977}, consisting of 34 nodes and 78 edges. We consider all the cliques as simplices, though not all the cliques are likely to be true higher-order interactions \cite{Krishnagopal2021}. According to Zachary's study \cite{Zachary1977}, the nodes are clustered into two communities: the \textit{Mr. Hi} community around node 1 and the \textit{Officer} community around node 34. We detect simplicial communities of Zachary's karate club network using our quantum algorithm. As shown in Figure \ref{fig:1}, $N_1=78$ edges (1-simplices) of Zachary's karate club network are assigned to two 1-down communities. Most elements of the first 1-down community (red edges in Fig. \ref{fig:1}) consist of the edges connected within the Mr. Hi group, whereas most elements of the second 1-down community (blue edges in Fig. \ref{fig:1}) consist of the edges connected within the Officer group. There are several edges (1-simplices) connecting the Officer group and the Mr. Hi group: (1, 32), (2, 31), (3, 10), (3, 28), and (3, 29) in the first 1-down community, and (3, 33), (9, 31), (9, 33), (9, 34), (14, 34), and (20, 34) in the second 1-down community. The simplicial modularity (Eq. \ref{simmodularity}) for the 1-down communities is $Q_1=0.434$.

We further detect higher-order simplicial communities of Zachary's karate club network. As highlighted in Table \ref{table:1}, $N_2=45$ triangles (2-simplices) are clustered into four 2-down communities, one of which is an isolated 2-simplex, (25, 26, 32). The simplicial modularity for the 2-down communities is $Q_2=0.515$. There are also three 3-down communities, two of which are isolated 3-simplices, (9, 31, 33, 34) and (24, 30, 33, 34). There is only one 4-down community consisting of (1, 2, 3, 4, 8) and (1, 2, 3, 4, 14) in the Mr. Hi group. The simplicial modularity values for the 3-down communities and 4-down community are zero because all the simplicial communities except one are isolated simplices.

\begin{table}[h!]
\caption{Simplicial communities of Zachary's karate club network.}
\label{table:1}
\centering
\begin{tabular}{llp{0.8\textwidth}}
\hline
\boldmath{$n$} & \boldmath{$N_n$} & \boldmath{$n$}\textbf{-down simplicial communities} \\ \hline
1 & 78 & (see Fig. \ref{fig:1}) \\ \hline
2 & 45 & \begin{itemize}
    \item (1, 2, 3), (1, 2, 4), (1, 2, 8), (1, 2, 14), (1, 2, 18), (1, 2, 20), (1, 2, 22), (1, 3, 4), (1, 3, 8), (1, 3, 9), (1, 3, 14), (1, 4, 8), (1, 4, 13), (1, 4, 14), (2, 3, 4), (2, 3, 8), (2, 3, 14), (2, 4, 8), (2, 4, 14), (3, 4, 8), (3, 4, 14)
    \item (3, 9, 33), (9, 31, 33), (9, 31, 34), (9, 33, 34), (15, 33, 34), (16, 33, 34), (19, 33, 34), (21, 33, 34), (23, 33, 34), (24, 28, 34), (24, 30, 33), (24, 30, 34), (24, 33, 34), (27, 30, 34), (29, 32, 34), (30, 33, 34), (31, 33, 34), (32, 33, 34)
    \item (1, 5, 7), (1, 5, 11), (1, 6, 7), (1, 6, 11), (6, 7, 17)
    \item (25, 26, 32)
\end{itemize} \\ \hline
3 & 11 & \begin{itemize}
    \item (1, 2, 3, 4), (1, 2, 3, 8), (1, 2, 3, 14), (1, 2, 4, 8), (1, 2, 4, 14), (1, 3, 4, 8), (1, 3, 4, 14), (2, 3, 4, 8), (2, 3, 4, 14)
    \item (9, 31, 33, 34)
    \item (24, 30, 33, 34)
\end{itemize} \\ \hline
4 & 2 & \begin{itemize}
    \item (1, 2, 3, 4, 8), (1, 2, 3, 4, 14)
\end{itemize} \\ \hline
\end{tabular}
\end{table}

Although the proposed quantum walk algorithm was only tested on Zachary's karate club network, we expect that similar quantum walk frameworks can be utilized to explore community structures in real networks, such as protein-protein interaction networks \cite{Saarinen2023}. It may be worth comparing the proposed quantum walk algorithm with classical methods, though there has been no direct classical random walk algorithm for simplicial community detection.

\section{Conclusion}
This study investigates discrete-time quantum walks on simplicial complexes, which represent a higher-order extension of graphs or networks. Since the conventional classical random walk cannot directly detect community structures, we present a discrete-time quantum walk algorithm for the identification of higher-order community structures called simplicial communities. The symmetry between up-communities and down-communities is presented. Our approach utilizes the Fourier coin to generate entangled translation states among neighboring simplices in a simplicial complex. The proposed quantum walk algorithm is tested in Zachary's karate club network as a test case; however, it needs to be validated through various examples and mathematical proof. This research has the potential to contribute to the understanding of complex systems at the crossroads of algebraic topology and quantum walk algorithms.

\section*{Acknowledgement}
This independent research received no external funding and was conducted by the author in 2023--2024. The author is a founder of TopolyTech in Korea. The author is grateful to Dr. U Jin Choi (Department of Mathematical Sciences, KAIST) for his advice and insightful discussions on persistent homology. The author would like to thank anonymous referees for their valuable comments.

\section*{Authorship contribution}
\textbf{Euijun Song:} Conceptualization, Methodology, Formal analysis, Investigation, Visualization, Writing – original draft.

\medskip

\appendix
\section{Appendix: Proof of the property of the long-time average} \label{appendix1}
We prove the following property of the long-time average (Eq. \ref{longtimeeigen}):
$$\lim_{T\to\infty} \frac{1}{T} \sum_{t=1}^{T} e^{i(\theta_{k_1}-\theta_{k_2})t} = \delta_{\theta_{k_1}, \theta_{k_2}}$$
where $i=\sqrt{-1}$ and $\theta_{k_1}, \theta_{k_2}\in [0, 2\pi)$.

\begin{proof}
(i) If $\theta_{k_1}=\theta_{k_2}$,
$$\lim_{T\to\infty} \frac{1}{T} \sum_{t=1}^{T} e^{i(\theta_{k_1}-\theta_{k_2})t} = \lim_{T\to\infty} \frac{1}{T} \sum_{t=1}^{T} 1 = 1.$$
(ii) If $\theta_{k_1}\neq\theta_{k_2}$,
\begin{eqnarray*}
\lim_{T\to\infty} \frac{1}{T} \sum_{t=1}^{T} e^{i(\theta_{k_1}-\theta_{k_2})t} &=& \lim_{T\to\infty} \frac{1}{T} \cdot \frac{e^{i(\theta_{k_1}-\theta_{k_2})} \left(1-e^{i(\theta_{k_1}-\theta_{k_2})T} \right)}{1-e^{i(\theta_{k_1}-\theta_{k_2})}} \\
&=& \frac{e^{i(\theta_{k_1}-\theta_{k_2})}}{1-e^{i(\theta_{k_1}-\theta_{k_2})}} \lim_{T\to\infty} \frac{1-e^{i(\theta_{k_1}-\theta_{k_2})T}}{T} \\
&=& 0.
\end{eqnarray*}
The last line holds because $\left| 1-e^{i(\theta_{k_1}-\theta_{k_2})T}\right| \leq2$.
\end{proof}

\section{Appendix: Error analysis of the finite-time average} \label{appendix2}
We prove the following error term of the finite-time average (Eq. \ref{errorterm}):
$$\left| \tilde{q}_{T}(\sigma_x\to\sigma_y) - q(\sigma_x\to\sigma_y) \right| = O\left(\frac{1}{T}\right).$$

\begin{proof}
According to the proof in Appendix \ref{appendix1}, we can obtain the following finite-time version of Eq. \ref{longtimeeigen}:
$$\frac{1}{T} \sum_{t=1}^{T} e^{i(\theta_{k_1}-\theta_{k_2})t} =
\begin{cases}
1 & \text{if } \theta_{k_1}=\theta_{k_2} \\
O(1/T) & \text{if } \theta_{k_1}\neq\theta_{k_2}.
\end{cases}$$
Hence, the finite-time version of Eq. \ref{transitionprob2} is given by
$$\frac{1}{T} \sum_{t=1}^{T} \left| \langle\sigma_y\to\sigma_w \vert U^{t} \vert \sigma_x\to\sigma_v\rangle \right|^2
= \sum_{k=1}^{m_n} \left| \langle\sigma_y\to\sigma_w \vert \Phi_{k}\rangle \langle\Phi_{k} \vert \sigma_x\to\sigma_v\rangle \right|^2 + O\left(\frac{1}{T}\right).$$
By combining this result with Eq. \ref{transitionprob1} and Eq. \ref{probtheory}, we have
$$\left| \tilde{q}_{T}(\sigma_x\to\sigma_y) - q(\sigma_x\to\sigma_y) \right| = O\left(\frac{1}{T}\right).$$
\end{proof}

\medskip

\bibliography{refs}

\begin{thebibliography}{31}
\providecommand{\natexlab}[1]{#1}
\providecommand{\url}[1]{\texttt{#1}}
\expandafter\ifx\csname urlstyle\endcsname\relax
  \providecommand{\doi}[1]{doi: #1}\else
  \providecommand{\doi}{doi: \begingroup \urlstyle{rm}\Url}\fi

\bibitem[Aharonov et~al.(1993)Aharonov, Davidovich, and Zagury]{Aharonov1993}
Y.~Aharonov, L.~Davidovich, and N.~Zagury.
\newblock Quantum random walks.
\newblock \emph{Phys. Rev. A}, 48:\penalty0 1687--1690, 1993.
\newblock \doi{10.1103/PhysRevA.48.1687}.

\bibitem[Travaglione and Milburn(2002)]{Travaglione2002}
B.~C. Travaglione and G.~J. Milburn.
\newblock Implementing the quantum random walk.
\newblock \emph{Phys. Rev. A}, 65:\penalty0 032310, 2002.
\newblock \doi{10.1103/PhysRevA.65.032310}.

\bibitem[Konno(2002)]{Konno2002}
Norio Konno.
\newblock Quantum random walks in one dimension.
\newblock \emph{Quantum Information Processing}, 1\penalty0 (5):\penalty0
  345--354, 2002.
\newblock \doi{10.1023/A:1023413713008}.

\bibitem[Konno(2005)]{Konno2005}
Norio Konno.
\newblock Limit theorem for continuous-time quantum walk on the line.
\newblock \emph{Phys. Rev. E}, 72:\penalty0 026113, 2005.
\newblock \doi{10.1103/PhysRevE.72.026113}.

\bibitem[Berry and Wang(2010)]{Berry2010}
Scott~D. Berry and Jingbo~B. Wang.
\newblock Quantum-walk-based search and centrality.
\newblock \emph{Phys. Rev. A}, 82:\penalty0 042333, 2010.
\newblock \doi{10.1103/PhysRevA.82.042333}.

\bibitem[Grover(1996)]{Grover1996}
Lov~K. Grover.
\newblock A fast quantum mechanical algorithm for database search.
\newblock In \emph{Proceedings of the Twenty-Eighth Annual ACM Symposium on
  Theory of Computing}, STOC '96, pages 212--219. Association for Computing
  Machinery, 1996.
\newblock \doi{10.1145/237814.237866}.

\bibitem[Mukai and Hatano(2020)]{Mukai2020}
Kanae Mukai and Naomichi Hatano.
\newblock Discrete-time quantum walk on complex networks for community
  detection.
\newblock \emph{Phys. Rev. Res.}, 2:\penalty0 023378, 2020.
\newblock \doi{10.1103/PhysRevResearch.2.023378}.

\bibitem[Matsue et~al.(2016)Matsue, Ogurisu, and Segawa]{Matsue2016}
Kaname Matsue, Osamu Ogurisu, and Etsuo Segawa.
\newblock Quantum walks on simplicial complexes.
\newblock \emph{Quantum Information Processing}, 15\penalty0 (5):\penalty0
  1865--1896, 2016.
\newblock \doi{10.1007/s11128-016-1247-6}.

\bibitem[Matsue et~al.(2018)Matsue, Ogurisu, and Segawa]{Matsue2018}
Kaname Matsue, Osamu Ogurisu, and Etsuo Segawa.
\newblock Quantum search on simplicial complexes.
\newblock \emph{Quantum Studies: Mathematics and Foundations}, 5\penalty0
  (4):\penalty0 551--577, 2018.
\newblock \doi{10.1007/s40509-017-0144-8}.

\bibitem[Giusti et~al.(2016)Giusti, Ghrist, and Bassett]{Giusti2016}
Chad Giusti, Robert Ghrist, and Danielle~S. Bassett.
\newblock Two's company, three (or more) is a simplex.
\newblock \emph{Journal of Computational Neuroscience}, 41\penalty0
  (1):\penalty0 1--14, 2016.
\newblock \doi{10.1007/s10827-016-0608-6}.

\bibitem[Hatcher(2002)]{Hatcher2002}
Allen Hatcher.
\newblock \emph{Algebraic topology}.
\newblock Cambridge University Press, 2002.

\bibitem[Jonsson(2008)]{Jonsson2008}
Jakob Jonsson.
\newblock \emph{Simplicial complexes of graphs}, volume~3.
\newblock Springer, 2008.

\bibitem[Mill\'an et~al.(2020)Mill\'an, Torres, and Bianconi]{Millan2020}
Ana~P. Mill\'an, Joaqu\'{\i}n~J. Torres, and Ginestra Bianconi.
\newblock Explosive higher-order kuramoto dynamics on simplicial complexes.
\newblock \emph{Phys. Rev. Lett.}, 124:\penalty0 218301, 2020.
\newblock \doi{10.1103/PhysRevLett.124.218301}.

\bibitem[Song(2023)]{Song2023}
Euijun Song.
\newblock Persistent homology analysis of type 2 diabetes genome-wide
  association studies in protein--protein interaction networks.
\newblock \emph{Frontiers in Genetics}, 14:\penalty0 1270185, 2023.
\newblock \doi{10.3389/fgene.2023.1270185}.

\bibitem[Vipond et~al.(2021)Vipond, Bull, Macklin, Tillmann, Pugh, Byrne, and
  Harrington]{Vipond2021}
Oliver Vipond, Joshua~A. Bull, Philip~S. Macklin, Ulrike Tillmann,
  Christopher~W. Pugh, Helen~M. Byrne, and Heather~A. Harrington.
\newblock Multiparameter persistent homology landscapes identify immune cell
  spatial patterns in tumors.
\newblock \emph{Proceedings of the National Academy of Sciences}, 118\penalty0
  (41):\penalty0 e2102166118, 2021.
\newblock \doi{10.1073/pnas.2102166118}.

\bibitem[Girvan and Newman(2002)]{Girvan2002}
M.~Girvan and M.~E.~J. Newman.
\newblock Community structure in social and biological networks.
\newblock \emph{Proceedings of the National Academy of Sciences}, 99\penalty0
  (12):\penalty0 7821--7826, 2002.
\newblock \doi{10.1073/pnas.122653799}.

\bibitem[Newman(2006)]{Newman2006}
M.~E.~J. Newman.
\newblock Modularity and community structure in networks.
\newblock \emph{Proceedings of the National Academy of Sciences}, 103\penalty0
  (23):\penalty0 8577--8582, 2006.
\newblock \doi{10.1073/pnas.0601602103}.

\bibitem[Mitzenmacher and Upfal(2005)]{Mitzenmacher2005}
Michael Mitzenmacher and Eli Upfal.
\newblock \emph{Probability and computing: Randomization and probabilistic
  techniques in algorithms and data analysis}.
\newblock Cambridge University Press, 2005.

\bibitem[Faccin et~al.(2014)Faccin, Migda\l{}, Johnson, Bergholm, and
  Biamonte]{Faccin2014}
Mauro Faccin, Piotr Migda\l{}, Tomi~H. Johnson, Ville Bergholm, and Jacob~D.
  Biamonte.
\newblock Community detection in quantum complex networks.
\newblock \emph{Phys. Rev. X}, 4:\penalty0 041012, 2014.
\newblock \doi{10.1103/PhysRevX.4.041012}.

\bibitem[Newman(2013)]{Newman2013}
M.~E.~J. Newman.
\newblock Spectral methods for community detection and graph partitioning.
\newblock \emph{Phys. Rev. E}, 88:\penalty0 042822, 2013.
\newblock \doi{10.1103/PhysRevE.88.042822}.

\bibitem[Krishnagopal and Bianconi(2021)]{Krishnagopal2021}
Sanjukta Krishnagopal and Ginestra Bianconi.
\newblock Spectral detection of simplicial communities via hodge laplacians.
\newblock \emph{Phys. Rev. E}, 104:\penalty0 064303, 2021.
\newblock \doi{10.1103/PhysRevE.104.064303}.

\bibitem[Mackay et~al.(2002)Mackay, Bartlett, Stephenson, and
  Sanders]{Mackay2002}
T.~D. Mackay, S.~D. Bartlett, L.~T. Stephenson, and B.~C. Sanders.
\newblock Quantum walks in higher dimensions.
\newblock \emph{Journal of Physics A: Mathematical and General}, 35\penalty0
  (12):\penalty0 2745, 2002.
\newblock \doi{10.1088/0305-4470/35/12/304}.

\bibitem[Saito(2019)]{Saito2019}
Kei Saito.
\newblock Periodicity for the fourier quantum walk on regular graphs.
\newblock \emph{Quantum Info. Comput.}, 19\penalty0 (1-2):\penalty0 23--34,
  2019.
\newblock \doi{10.26421/QIC19.1-2-3}.

\bibitem[Lim(2020)]{Lim2020}
Lek-Heng Lim.
\newblock Hodge laplacians on graphs.
\newblock \emph{SIAM Review}, 62\penalty0 (3):\penalty0 685--715, 2020.
\newblock \doi{10.1137/18M1223101}.

\bibitem[Baccini et~al.(2022)Baccini, Geraci, and Bianconi]{Baccini2022}
Federica Baccini, Filippo Geraci, and Ginestra Bianconi.
\newblock Weighted simplicial complexes and their representation power of
  higher-order network data and topology.
\newblock \emph{Phys. Rev. E}, 106:\penalty0 034319, 2022.
\newblock \doi{10.1103/PhysRevE.106.034319}.

\bibitem[Eckmann(1944)]{Eckmann1944}
Beno Eckmann.
\newblock Harmonische funktionen und randwertaufgaben in einem komplex.
\newblock \emph{Commentarii Mathematici Helvetici}, 17\penalty0 (1):\penalty0
  240--255, 1944.

\bibitem[Estrada and Ross(2018)]{Estrada2018}
Ernesto Estrada and Grant~J. Ross.
\newblock Centralities in simplicial complexes. applications to protein
  interaction networks.
\newblock \emph{Journal of Theoretical Biology}, 438:\penalty0 46--60, 2018.
\newblock \doi{10.1016/j.jtbi.2017.11.003}.

\bibitem[Hern{\'a}ndez~Serrano and S{\'a}nchez~G{\'o}mez(2020)]{Hernandez2020}
Daniel Hern{\'a}ndez~Serrano and Dar{\'\i}o S{\'a}nchez~G{\'o}mez.
\newblock Centrality measures in simplicial complexes: Applications of
  topological data analysis to network science.
\newblock \emph{Applied Mathematics and Computation}, 382:\penalty0 125331,
  2020.
\newblock \doi{10.1016/j.amc.2020.125331}.

\bibitem[Muhammad and Egerstedt(2006)]{Muhammad2006}
Abubakr Muhammad and Magnus Egerstedt.
\newblock Control using higher order laplacians in network topologies.
\newblock In \emph{Proc. of 17th International Symposium on Mathematical Theory
  of Networks and Systems}, pages 1024--1038, 2006.

\bibitem[Zachary(1977)]{Zachary1977}
Wayne~W. Zachary.
\newblock An information flow model for conflict and fission in small groups.
\newblock \emph{Journal of Anthropological Research}, 33\penalty0 (4):\penalty0
  452--473, 1977.
\newblock \doi{10.1086/jar.33.4.3629752}.

\bibitem[Saarinen et~al.(2023)Saarinen, Goldsmith, Wang, Loscalzo, and
  Maniscalco]{Saarinen2023}
Harto Saarinen, Mark Goldsmith, Rui-Sheng Wang, Joseph Loscalzo, and Sabrina
  Maniscalco.
\newblock Disease gene prioritization with quantum walks.
\newblock \emph{arXiv preprint}, 2311.05486, 2023.
\newblock \doi{10.48550/arXiv.2311.05486}.

\end{thebibliography}

\end{document}